\documentclass[journal,twocolumn]{IEEEtran}
\newlength{\doublecolumnwidth}
\usepackage{cite}

\ifCLASSINFOpdf
  \usepackage[pdftex]{graphicx}
\else
  \usepackage[dvips]{graphicx}
\fi
\usepackage[cmex10]{amsmath}
\usepackage{amsfonts}
\usepackage{color}
\def\eq#1{Eq.~\eqref{eq:#1}}
\def\fig#1{Fig.~\ref{fig:#1}}
\def\eq#1{Eq.~\eqref{eq:#1}}

\def\cE{\mathcal{E}}

\def\cU{\mathcal{U}}

\def\eq#1{Eq.~\eqref{eq:#1}}
\def\fig#1{Fig.~\ref{fig:#1}}
\def\sec#1{Sec.~\ref{sec:#1}}

\begin{document}
%
\title{Branching MERA codes: \\a natural extension of polar codes}
%
%
%

\author{Andrew~James~Ferris
        and~David~Poulin
\thanks{A. Ferris is with ICFO---Institut de Ciencies Fotoniques, Parc Mediterrani de la Tecnologia, 08860 Barcelona, Spain as well as Max-Planck-Institut f\"ur Quantenoptik, Hans-Kopfermann-Str. 1, 85748 Garching, Germany, and was recently with the Departement de Physique, l'Universit\'e de Sherbrooke,
Sherbrooke, Qu\'ebec J1K 2R1, Canada. e-mail: andy.ferris@icfo.es.}
\thanks{D. Poulin is with the Departement de Physique, l'Universit\'e de Sherbrooke,
Sherbrooke, Qu\'ebec J1K 2R1, Canada.}
\thanks{This work was funded by the National Research Council Canada.}}

\maketitle

\begin{abstract}
We introduce a new class of circuits for constructing efficiently decodable error-correction codes, based on a recently discovered contractible tensor network \cite{EV12b}. We perform an in-depth study of a particular example that can be thought of as an extension to Arikan's polar code \cite{A09a}. Notably, our numerical simulation show that this code polarizes the logical channels more strongly while retaining the log-linear decoding complexity using the successive cancellation decoder. These codes also display improved error-correcting capability with only a minor impact on decoding complexity. Efficient decoding is realized using powerful graphical calculus tools developed in the field of quantum many-body physics. In a companion paper \cite{FP13}, we generalize our construction to the quantum setting and describe more in-depth the relation between classical and quantum error correction and the graphical calculus of tensor networks.
\end{abstract}


%
\IEEEpeerreviewmaketitle

\section{Introduction}
%
%
%
%

The phenomenon of channel polarization, discovered by Arikan \cite{A09a}, can be produced by a controlled-not (CNOT) gate. Because the control bit is added to the target bit, it becomes redundantly encoded and thus effectively more robust. On the other hand, the information of the target bit is partially washed away because its value is  modified in a way that depends on the  value of the possibly unknown control bit. We thus say that the channels have partially polarized into a better and a worse channel. The encoding circuit of a polar code is obtained by recursing this polarization procedure, and asymptotically produces a perfect polarization, where  a fraction of the channels are error-free and the complement are completely randomizing. 

Because of this recursive nature, the encoding circuit takes the geometric form of a spectral transformation where CNOT gates follow a hierarchical arrangement on different length-scales (depicted in \fig{circuits} (a)), and much like the (fast) Fourier transform, the linear encoding matrix can be decomposed into a Kronecker product of small matrices.  In this case, the polarization is defined with respect to the successive cancellation decoder, where the marginal probability of input bit $i$ is calculated with the prior knowledge of the bits $1,\dots,i$. Using this construction, Arikan was able to give the first concrete example of a provably efficient and capacity-achieving code (for symmetric channels) \cite{A09a}, generating significant interest and stimulating further work on improvements and generalizations, e.g., \cite{AT09a,STA09a,KU10a,KSU10a,MV11a}.  In this Article, we present a natural extension of polar code based on recently discovered contractible tensor networks \cite{EV12b}.

Abstractly, we can view a gate, such as a CNOT, as a tensor $A^{\alpha \beta\gamma\ldots}$ with a certain number of indices denoted $\alpha, \beta, \gamma,\ldots$, each taking values in a finite set, that we will assume henceforth to be $\mathbb{Z}_2$.  The number of indices is the rank of the tensor. For instance, the CNOT gate is a rank-four tensor $N^{\alpha\beta\gamma \delta}$ with  indices $\alpha$ and $\beta$ representing the two input bits and $\gamma$ and $\delta$ representing the two output bits, and  the value of the tensor  given by $N^{\alpha\beta\gamma\delta} = 1$ if $\gamma = \alpha$ and $\delta = \alpha\oplus\beta$, and $N^{\alpha\beta\gamma\delta} = 0$ otherwise. We can graphically represent a tensor as a vertex and its indices as edges, with the degree of the vertex equal to the rank of the tensor. In that setting, an edge linking two vertices represents a tensor contraction defined by the following equation
\begin{equation}\includegraphics[width=6.5cm]{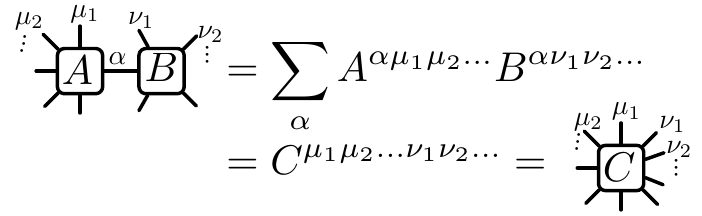}.\label{eq:TNC}\end{equation}
It follows from this definition that a graph represents a TN with all edges contracted, and hence a scalar. 

\begin{figure}[t]
\centering
\includegraphics[width=\doublecolumnwidth]{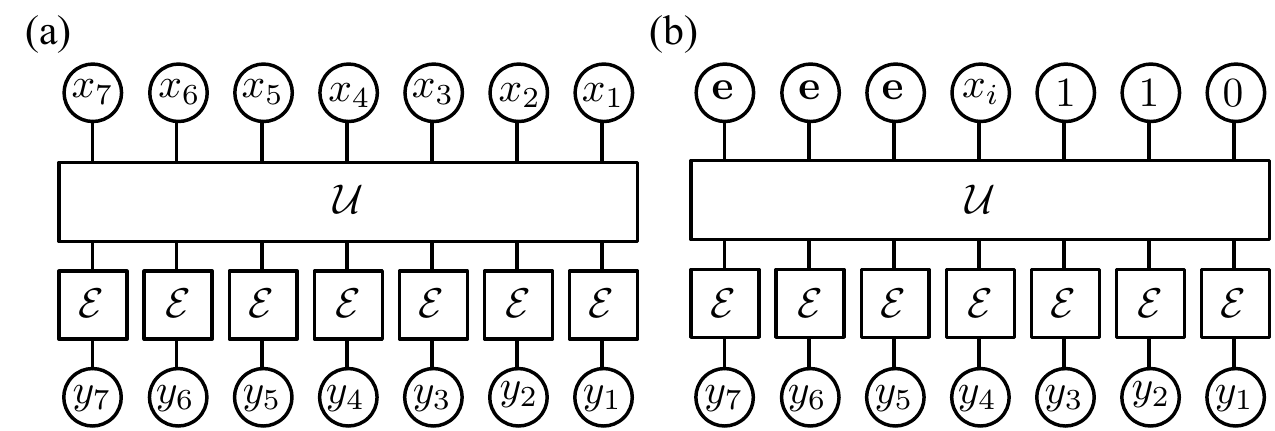}
\caption{(a) A simple TN diagram of the generic decoding problem. The $n$ input bits $x_i$ are a combination of $k$ data bits and $n-k$ frozen bits, which are passed through the encoding circuit $\mathcal{U}$. Given the measurements $y_i$ and the symmetric noise model $\cE$, we wish to determine the most likely configuration of data bits. The unnormalized probability $P(\mathbf{x}|\mathbf{y})$ is given by contracting the above TN, but it is not feasible to repeat for all $2^k$ possible inputs. (b) The successive cancellation decoder iteratively determines input bits in a right-to-left order. To determine the relative probabilities of bit $i$, we freeze the bits to the right using prior knowledge, while remaining completely ignorant about the states to the left, where $``{\bf e}"$ represents the uniform mixture $(1,1)$.  \label{fig:decoder}}
\end{figure}

Viewing the encoding circuit of a code --- such as a polar code encoding circuit shown at \fig{circuits} (a) --- as a TN enables us to recast the decoding problem as a TN contraction problem. An  encoding circuit $\cU$ is a rank-$2n$ tensor, with $n$ indices representing $n$ input bits and $n$ indices representing $n$ output bits, where some of the input bits are fixed (frozen) to 0. A single bit channel $\cE$ is a stochastic matrix, and hence a rank-two tensor. Finally, we can represent a bit as a rank-one tensor, with the tensor $``0" = (0,1)$ representing the bit value 0 and tensor  $``1" = (0,1)$ representing the bit value 1. Given these, the probability of the input bit string  $\mathrm{x} = (x_1,\dots,x_n)$ given the observed output $\mathrm{y} = (y_1,\dots,y_n)$ can be represented as the TN shown at \fig{decoder}~(a). 

In general, not all TNs can be efficiently contracted. Refering to \eq{TNC} where tensor $A$ has rank 6 and tensor $B$ has rank $5$, we see that the tensor resulting from their contraction has rank $6+5-2 = 9$. Thus, while tensor $A$ is specified by $2^6$ entries and tensor $B$ is specified by $2^5$ entries, tensor $C$ contains $2^9 \gg 2^6+2^5$ entries. Thus, a TN composed of bounded-rank tensors (e.g., a circuit with only two-bit gates) can be specified efficiently. However, the tensors obtained at intermediate steps of the TN contraction schedule can be of very high rank $r$, and so its contraction will produce an intractable amount of data $2^r$. The contraction schedule that minimizes the intermediate tensor rank defines the tree-width of the graph, so generally the cost of contracting a TN is exponential with its tree-width \cite{MS08a}. 

This implies that encoding circuits that produce TNs with finite tree-width can be efficiently decoded. This is the case for instance of convolutional codes \cite{JZ99a}, whose corresponding TN is simply a chain, and therefore have constant tree-width. However, it can sometimes be possible to contract TNs with large tree-width by making use of special circuit identities. An example is provided by the fact that a CNOT gate with a 0 entry on the controlled bit is equivalent to the identity, c.f. \fig{identities}~(b) for the corresponding graphical identity. The combination of such circuit identities provide a powerful graphical calculus that can be used to contract highly complex TNs. In particular, Arikan's sequential cancellation decoding can be recast in this graphical calculus as an efficient TN contraction, see \sec{C-decoding}. 

Graphical calculus is commonly used in quantum physics, starting with Feynman diagrams for quantum electrodynamics, to quantum circuit representation of quantum computations. More recently, a graphical calculus was developed for the representation of quantum many-body states \cite{V03a,VC04a,SDV06a,Vid05a,EV12b}. The quantum state of a system comprising $n$ particles is a $2^n$-dimensional vector, so its specification requires an exponential amount of data. A vector with $2^n$ component can be viewed as  a rank-$n$ tensor with binary indices. Thus, by restricting  to  tensors that are obtained from the contraction of polynomially many bounded-degree tensors, we reduce the amount of data required to specify a quantum state from exponential to polynomial. Then, the evaluation of physical quantities of interest (energy, magnetization, etc.) amounts to the problem of contracting the corresponding tensor network. 

Tensor networks therefore establish a relation between error correction and quantum many-body physics. Here, we use this relation to propose a new classical coding scheme that builds on the powerful graphical calculus developed in the field of quantum many-body physics. In this field, the tensor network associated to polar codes are a restricted form of {\em branching multi-scale entanglement renormalization ansatz} (branching MERA) tensor networks~\cite{EV12b}. More precisely, they correspond to branching MERA networks with half of the tensors being trivial, resulting in an object that could be called a branching tree. By reinserting the missing tensors in the network, we obtain a new family of codes that we call {\em branching MERA codes}. These codes are a natural generalization of polar codes, and inherit many of their properties including a successive cancellation decoder that  produces a tensor contractible in log-linear time.  To demonstrate the power of this class, we compare the polar code to the next-simplest branching MERA code with twice as many CNOT gates. While the decoding algorithm is slower by a small, constant numerical factor, we observe a significant improvement in both the channel polarization and the error-correction performance. While an important practical limitation of polar code is their important finite-size effects \cite{KU10b}, we observe that branching MERA codes display a steeper waterfall region, thus suppressing such finite-size effects.


\subsection{Paper outline}
This paper is structured as follows. In the next section we detail the encoding circuit for our new code, comparing it to the polar code, and describe how to implement the successive-cancellation decoder. In Sec.~\ref{sec:matrix}, we describe some properties of the linear encoding matrix shared between both the polar and branching MERA codes, and why we believe these lead to enhanced error-correction performances. The channel polarization under successive cancellation of both codes are compared in Sec.~\ref{sec:polarization} followed by numerical results on the error-correction properties in Sec.~\ref{sec:results}. We conclude and present some future directions in Sec.~\ref{sec:discussion}.

\section{Encoding and decoding}
\label{sec:geo}

In this Section we  describe the encoding circuit of branching MERA codes, focusing on its relationship to Arikan's polar code, followed by an algorithm to implement successive cancellation decoding.

\subsection{Encoding circuit}

\begin{figure}[t]
\centering
\includegraphics[width=0.6\doublecolumnwidth]{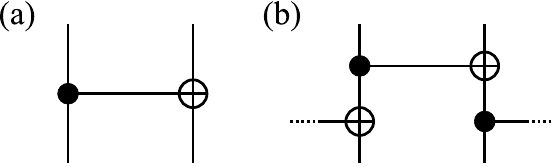}
\caption{The basic building blocks of (a) the polar code, and (b) the branching MERA code. The polar code can be thought of as an attempt to strengthen then left logical channel at the expense of the right, under successive cancellation decoding. In the branching MERA code, information from all input channels is spread out, from left to right and at different length scales (with periodic boundaries). A specific choice of successive decoding order  makes the left-most channels less susceptible to noise than the right, on average. \label{fig:primitives} }
\end{figure}

The polar code is based on the idea of polarization of the input channels into those that are almost noiseless or very noisy, under the successive cancellation decoder. The basic primitive of the polar code is the CNOT gate, as depicted in Fig.~\ref{fig:primitives}~(a). The CNOT copies the data from the left input (logical) channel and adds it to the right, essentially giving that data more opportunities to avoid corruption by the noise. Conversely, the data on the right logical channel may become obscured by uncertainty in the data from the left --- specifically because the successive cancellation decoder will not make that determination until later.

In our work, we go beyond this picture by continuing to copy the data in a left-to-right fashion by using a second layer of CNOT gates. The primitive in Fig.~\ref{fig:primitives}~(b) is applied everywhere on the lattice, connecting all sites (in a periodic structure) so that every logical channel becomes the control bit of at least one CNOT. The goal is to more evenly and rapidly spread out the information, in the hope that the data is better protected from noise, as well as to increase the impulsive response of the encoder (and thus possibly the distance of the code).

In both cases, these primitives are composed (or concatenated) on different length scales --- with the distance spanned by the CNOT doubling at each `layer' of the code. In the original polar code, it was observed that the channel polarization increased with each additional layer, with the large codes approaching the capacity of binary symmetric channels. With the branching MERA code, we continue to compose layers together, spreading out the information on exponentially growing length scales.

\begin{figure}[t]
\centering
\includegraphics[width=1\doublecolumnwidth]{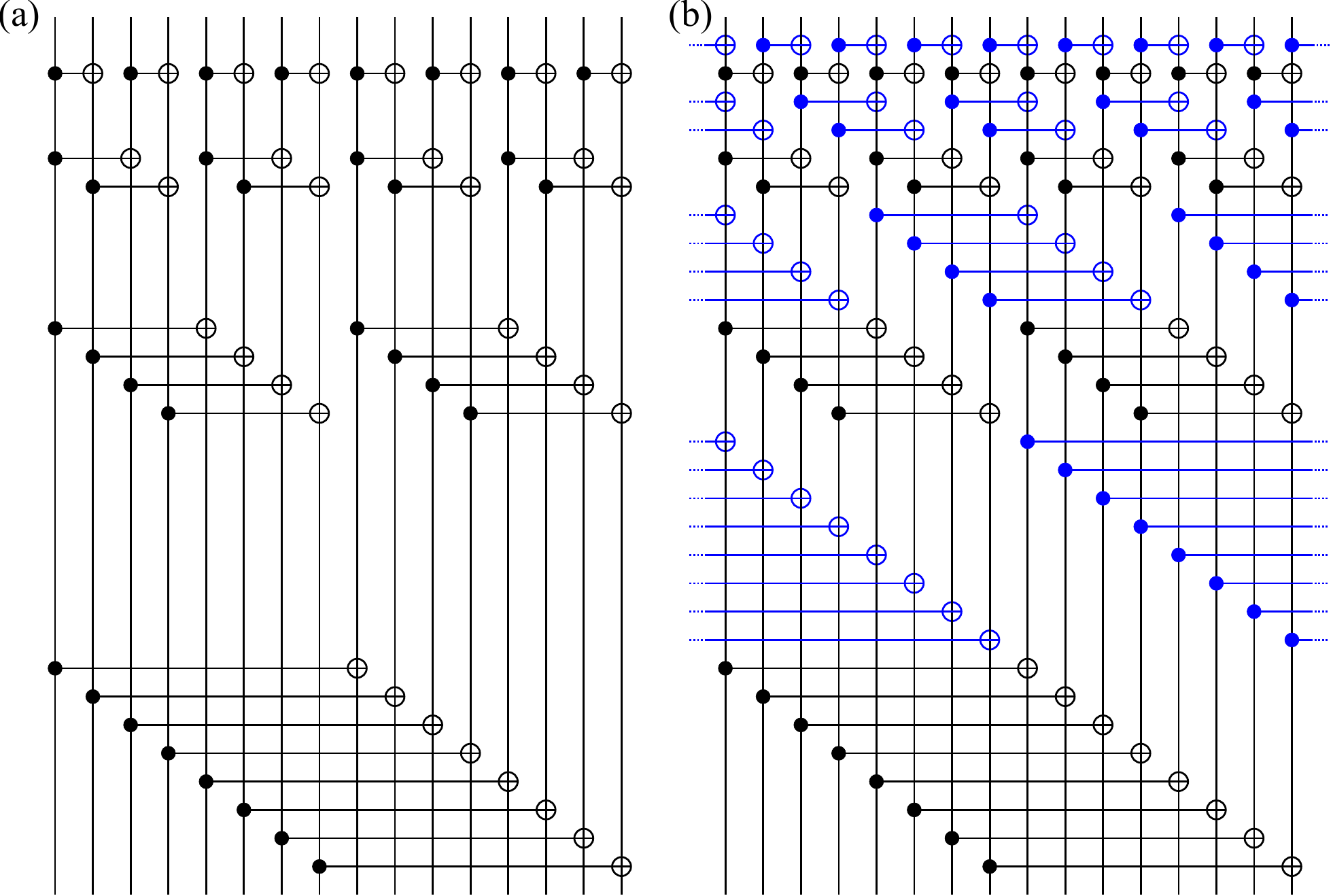}
\caption{The encoding circuits of (a) the polar code, and (b) the branching MERA code for $2^4 = 16$ sites. The polar code contains half of the gates of the branching MERA. The extra gates are highlighted in blue in (b), while the dots represent periodic boundaries. Branching MERA codes can be defined with our without periodic boundary conditions, the difference being the presence or absence of these dotted gates. \label{fig:circuits} }
\end{figure}

An example of the two encoding circuits with 16 bits (4 layers) is shown in Fig.~\ref{fig:circuits}. For $n$ bits, the polar code contains $n \log_2 n / 2$ CNOT gates, while the branching MERA code contains twice as many. The branching MERA code includes all of the gates of the polar code in addition to an alternating layer of gates, highlighted in blue. The depth of the circuit increases from $\log_2 n$ to $2\log_2 n$. Encoding times for these circuits grows log-linearly with $n$, and thus they can be considered efficient encoders for all practical purposes.

\subsection{Decoding algorithm}
\label{sec:C-decoding}

It is not immediately obvious that the polar or branching MERA codes can be efficiently decoded. One of Arikan's achievements was to realize that the decoding problem simplifies significantly under a successive cancellation scheme. In this decoder, the goal is to determine a single bit at the time, moving from right-to-left, by assuming complete ignorance of the input bits to the left, and total confidence in the value of the input bits to the right (either because they are frozen in the code, or because we have decoded those bits already). In Fig.~\ref{fig:decoder}, we write down the central probability density calculations required for a generic decoder and the successive cancellation decoder in a simple \emph{tensor network diagram}.

A generic, optimal decoder will locate the codeword with maximal likelihood --- that is, the most probable input $\mathrm{x} = (x_1,\dots,x_n)$ given the observed output $\mathrm{y} = (y_1,\dots,y_n)$, error model $\cE$, and the set of frozen bits $F$:
\begin{equation}
   \max_{\mathrm{x}} P(\mathrm{x} | \mathrm{y}, \{x_k,  \forall k \in F\}).
\end{equation}
However, for many codes determining the most probable codeword exactly is a hard problem, and a range of (usually iterative) approximations are used. The successive cancellation decoder begins with the right-most non-frozen bit at position $i$, and determines its value by maximizing
\begin{equation}
   \max_{x_i} P(x_i | \mathrm{y}, \{x_k,  k = i+1,\dots,n \}).
\end{equation}
For the purpose of the above calculation, the bits to the left of $i$ (i.e. $1,\dots,i-1$) are considered unknown, even if they are frozen bits. In this sense, successive cancellation is not an optimal decoder, because it does not take advantage of all the available information at every step. It then proceeds to the next non-frozen bit, and so on, until the complete message has been determined.

The tensor networks depicted in Fig.~\ref{fig:decoder} are calculations on bit probability distributions. There, we have used the tensor notation  $``0" = (0,1)$ and   $``1" = (0,1)$ as above, and have introduced the (un-normalized) uniform distribution ${\bf e} = (1,1)$ to represent a bit of which we have no knowledge. Furthermore, the measurement $y_i$ and the channel $\cE$ implies a probability distribution on the output bit $p_i$ that is given by Bayes' rule. For symmetric channels $\cE$ and linear encoding circuits this is simply $p_i \propto \cE^T y_i$.

\begin{figure}[t]
\centering
\includegraphics[width=1\doublecolumnwidth]{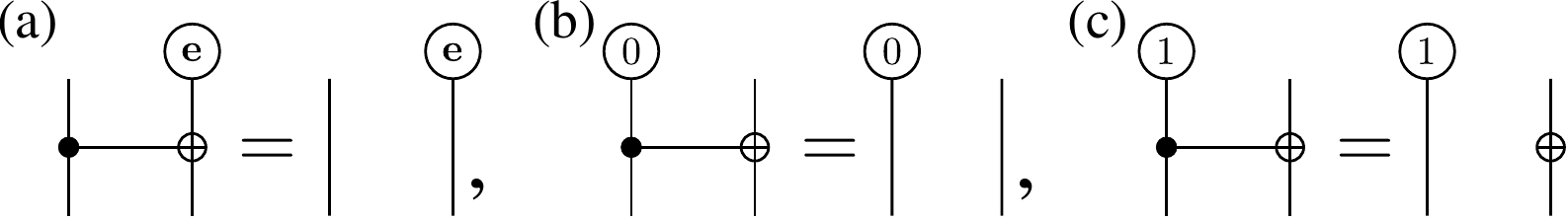}
\caption{Three basic circuit identities relating how the CNOT acts on probability distributions. These represent \emph{every} product input distribution that results in an uncorrelated, product output distribution. We must use these to simplify the tensor network contraction required for the successive cancellation decoder, for both the polar and branching MERA codes.  \label{fig:identities} }
\end{figure}

The CNOT gate has a very simple action on some states, such that it does not introduce any correlations to the joint-distribution. In Fig.~\ref{fig:identities} we detail {\em circuit identities} that define the action of CNOT on the distributions $``0"$, $``1"$, and $\bf e$. Generically, similar identities hold for all reversible (i.e. deterministic, one-to-one) gates.

\begin{figure}[t]
\centering
\includegraphics[width=1\doublecolumnwidth]{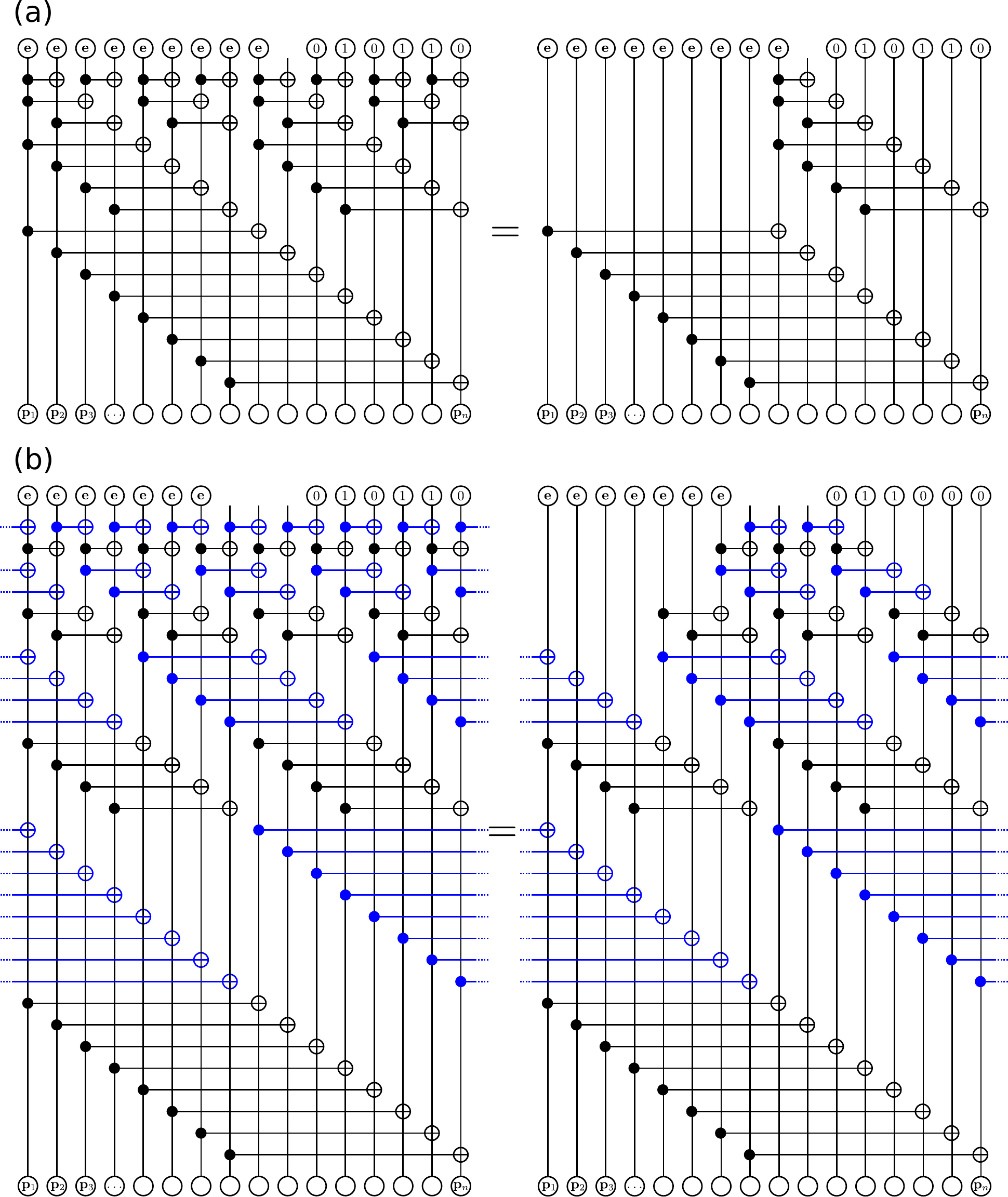}
\caption{The simplified tensor networks for successive cancellation decoding of (a) the polar code and (b) the branching MERA code, using the identities illustrated in Fig.~\ref{fig:identities}. These tensor networks contain open (non-contracted) indices as proxies for any tensor that could be placed at these location. For the branching MERA, it is natural to determine the joint probability distribution of three neighboring bits, because of the iterative scheme in Fig.~\ref{fig:contractions}~(b). Both tensor networks can be contracted from the bottom-up in a time \emph{linear} in the total number of bits, $n$. \label{fig:decoders} }
\end{figure}

Applying these identities to the polar and branching MERA codes results in a vast simplification. In fact, \emph{most} of  the CNOT gates are removed, and the number of remaining gates drops from $\mathcal{O}(n \log n)$ to $\mathcal{O}(n)$. This is illustrated at \fig{decoders} (a) for the polar code and (b) for the branching MERA code.  In both cases, the remaining tensor has a constant tree-width, and so it can be contracted efficiently. 

\begin{figure}[t]
\centering
\includegraphics[width=\doublecolumnwidth]{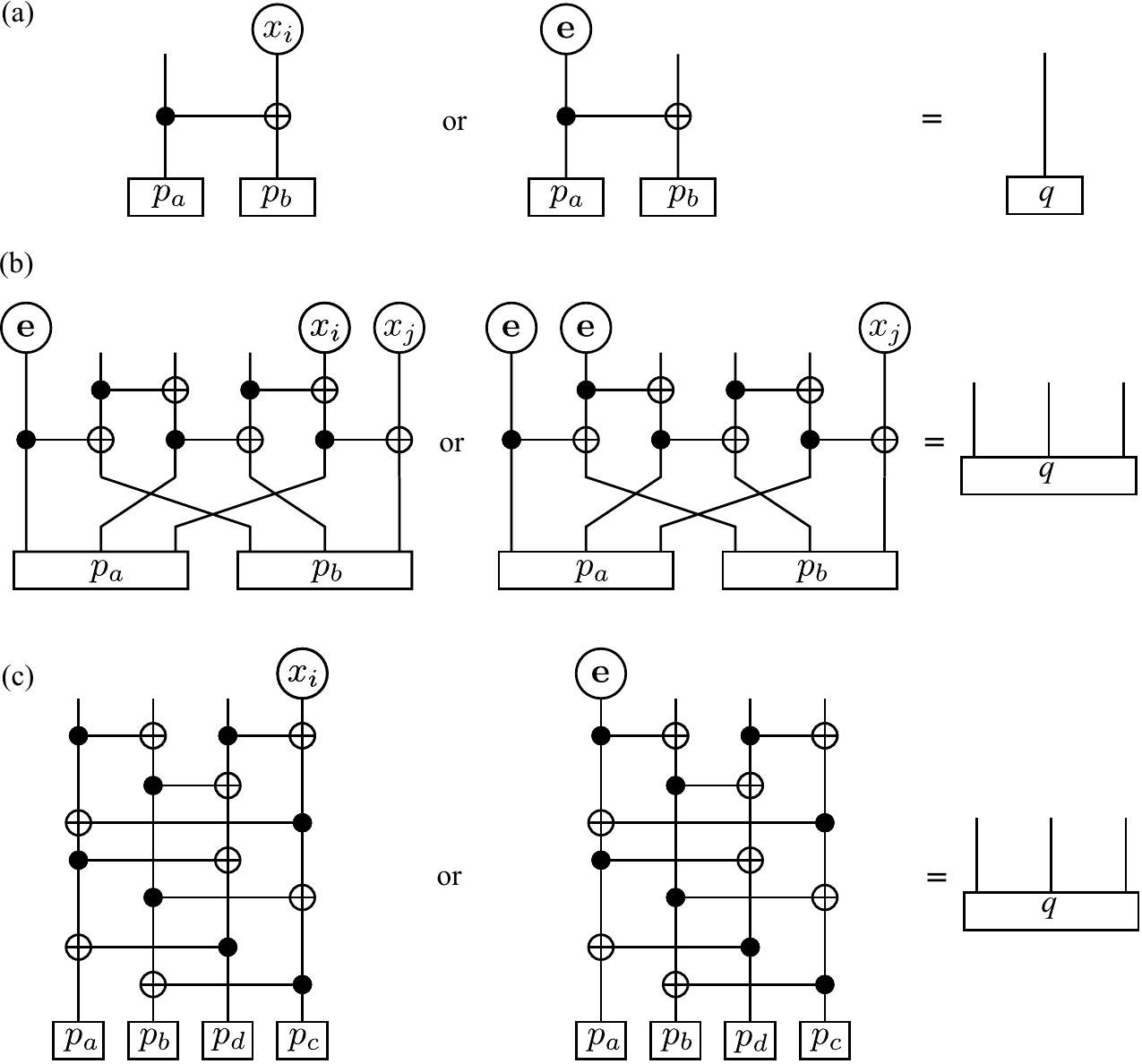}
\caption{The basic contractions required to decode (a) the polar code and (b,c) the branching MERA code. As opposed to the identities of \fig{identities}, these contraction identities apply to arbitrary tensors. The tensor $q$ depends on the specific tensor network being considered.  One of these basic transformations is selected at each `layer' of the code, depending on the input bit(s) targeted. (b) For the branching MERA code, a 3-bit to 3-bit transformation is natural~\cite{EV12b}. (c) The bottom two layers of the branching MERA (i.e. 4 bits) are required to build up to the 3-bit distribution. \label{fig:contractions} }
\end{figure}

To see how this is done, it is useful to highlight a few {\em contraction identities} shown at \fig{contractions}. In contrast to the circuit identities of \fig{identities} the contraction identities of \fig{contractions} are purely graphical in that they do not depend on the nature of the gates being contracted but only on the underlying graph structure. For instance, the CNOTs used in these contraction identities could be replaced by any other rank-four tensor and preserve the identities. Applying these graphical identities repeatedly starting from the the bottom of the diagrams of \fig{decoders} and working our way up yields an efficient contraction schedule. This demonstrates that sequential cancellation decoding can be realized as an efficient tensor network contraction in both the polar and branching MERA code.

\section{Linear encoding matrix}
\label{sec:matrix}

Before moving on to the performance of the branching-MERA code, we will briefly analyze the linear encoding matrix corresponding to the encoding circuits described in \fig{circuits}.

It is well known that the encoding matrix $G_n$ for Arikan's polar code can be decomposed into a Kronecker-product of $2 \times 2$ matrices,
\begin{equation}
    G_{2^l} = \left[ \begin{array}{cc} 1 & 1 \\ 0 & 1 \end{array} \right]^{\otimes l}.
\end{equation}
It also follows from this definition that $G_n^2 = I$, that is, the matrix is idempotent (under binary arithmetic), and that in some sense  it doesn't matter what order the layers of CNOTs are applied (the operations commute).

\begin{figure}[t]
\centering
\includegraphics[width=0.33\doublecolumnwidth]{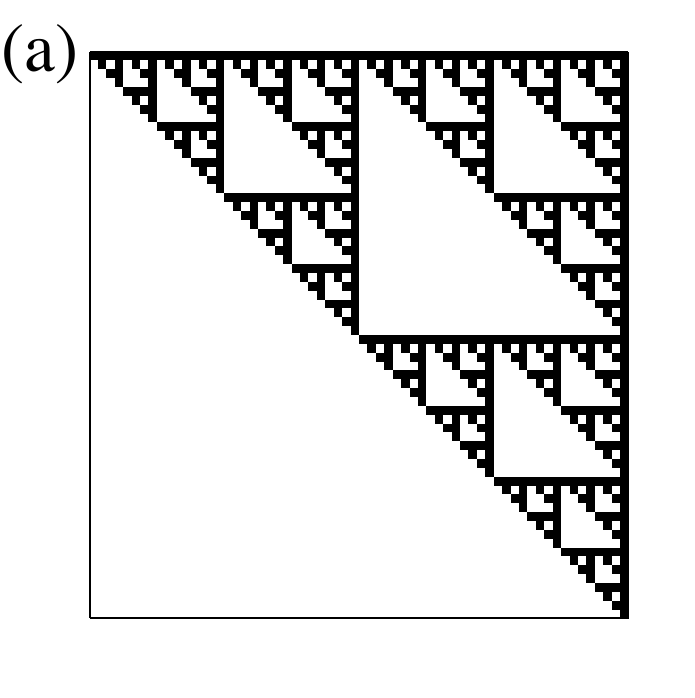}\includegraphics[width=0.33\doublecolumnwidth]{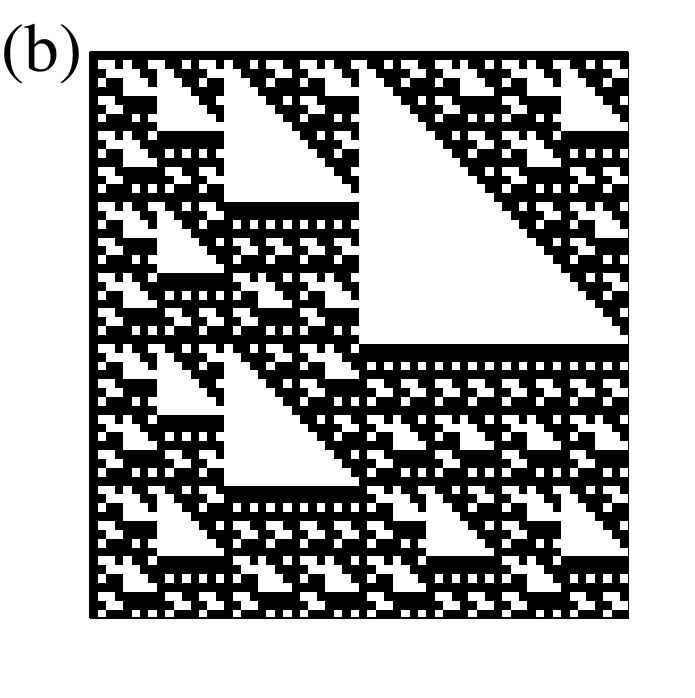}\includegraphics[width=0.33\doublecolumnwidth]{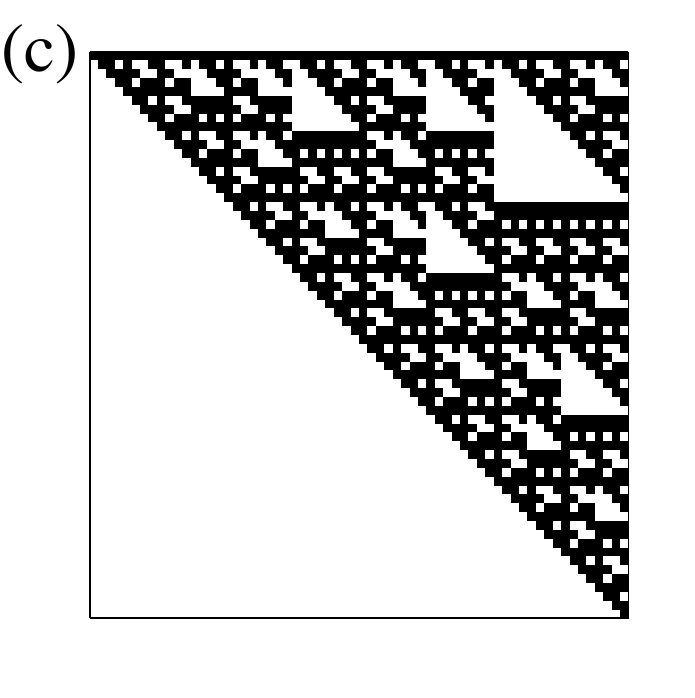}
\caption{Graphical representation of the linear encoding matrix for 64-bit codes for (a)~the Polar code, (b)~periodic branching-MERA code, and (c)~non-periodic branching-MERA code. Here white represents matrix entries of 0 and black represents 1. \label{fig:matrices} }
\end{figure}

On the other hand, the branching-MERA does not display these simple properties. In \fig{matrices} we display the contents of the encoding matrices for 64-bit polar and branching-MERA codes graphically. We know of no way of simplifying the description in terms of Kronecker-products or similar. The order of the layers in the encoding circuit will change the encoding matrix (and crucially, will affect the decoding complexity). While the encoding matrix is not idempotent (for codes larger than 2 bits), we have found numerically that $G_n^n = I$ under binary arithmetic. That is, applying the encoding circuit $n$ times to $n$ bits is equivalent to the identity operation.

\section{Channel polarization}
\label{sec:polarization}

The breakthrough achievement of the polar code was to prove, under a well-defined and efficient decoding scheme, that the logical channels corresponding to individual bits polarize into either `perfect' or `useless' channels in the limit of large codes --- and further that the ratio of good to bad channels tends to the (symmetric) capacity.

Nonetheless, there are noticeable finite-size effects in the polar code that make the performance less than ideal for practical code lengths. Here we study the channel polarization for both the polar and branching-MERA codes using the erasure channel as an example. We observe that channel polarization is stronger under the branching-MERA code resulting in reduction in the expected error rate. We obtain the same results for branching-MERA with and without periodic boundaries.  

Channel polarization occurs at the level of the induced logical channels that is the result of the encoding circuit and successive cancellation decoding. The $i$th logical channel is defined as the information obtained from channel $i$ given knowledge of the data sent on channels $i+1,\dots,n$, the received data, and the error model. In~\cite{A09a} it is shown that the CNOT gate acts to increase the amount of polarization between two channels, that is the left-channel becomes less noisy at the expense of the right. Furthermore, it was shown that two erasure channels with erasure rates $\epsilon_L$ and $\epsilon_R$ (on the left and right, respectively) transform into two new effective erasure channels with erasure rates $\epsilon_L^{\prime} = \epsilon_L \epsilon_R$ and $\epsilon_R^{\prime} = \epsilon_L + \epsilon_R - \epsilon_L \epsilon_R$. The transformation is slightly more complicated for the branching-MERA because correlated states of knowledge must be treated --- for example, we might know the sum of two bit values while not knowing either. We have implemented an exact calculation of the channel polarization of the branching-MERA code with the erasure channel. For other channels, we observe that Monte Carlo sampling can produce satisfactory results.

\begin{figure}[t]
\centering
\includegraphics[width=0.5\doublecolumnwidth]{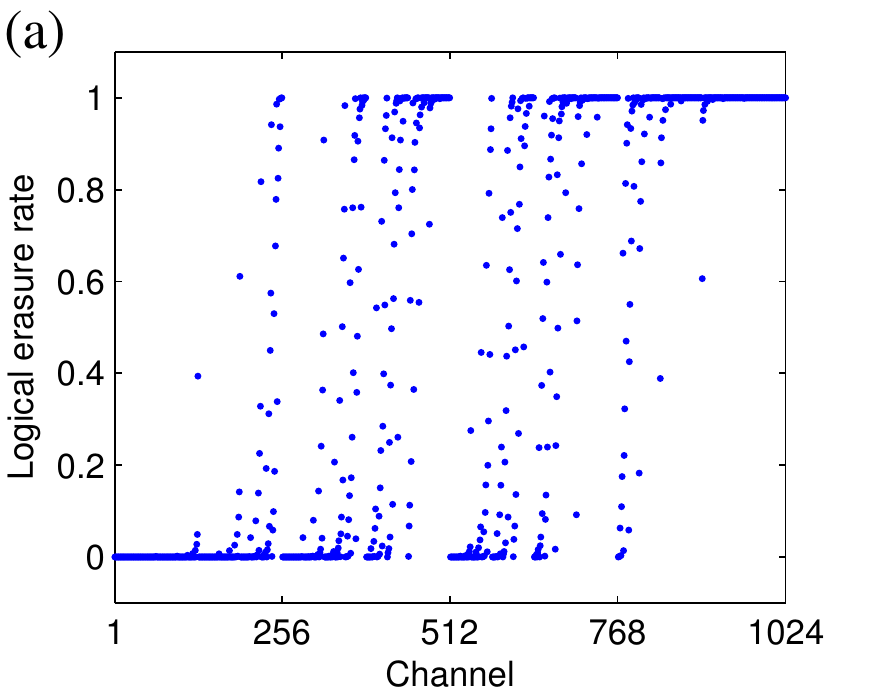}\includegraphics[width=0.5\doublecolumnwidth]{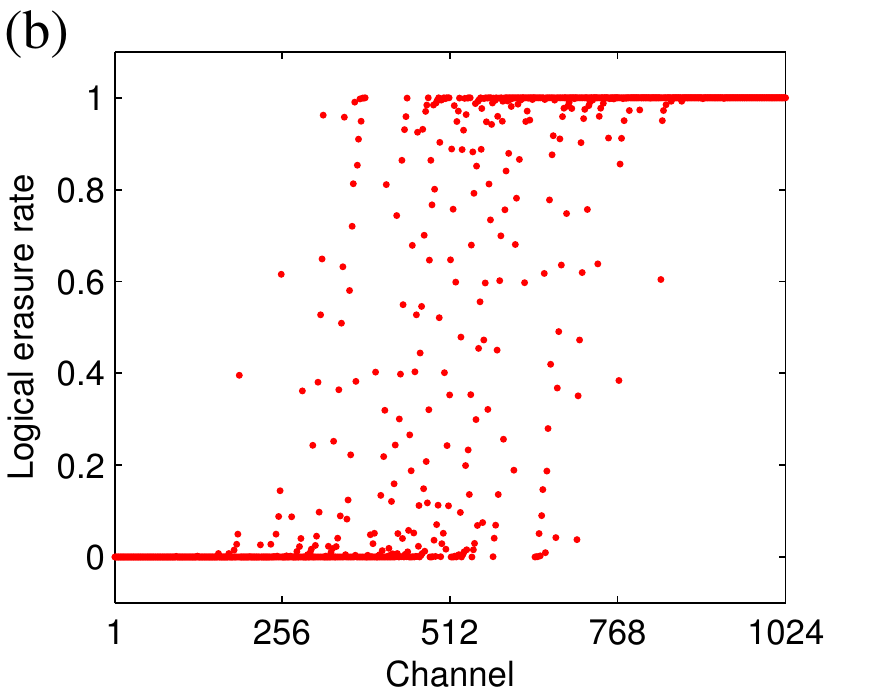}
\includegraphics[width=0.5\doublecolumnwidth]{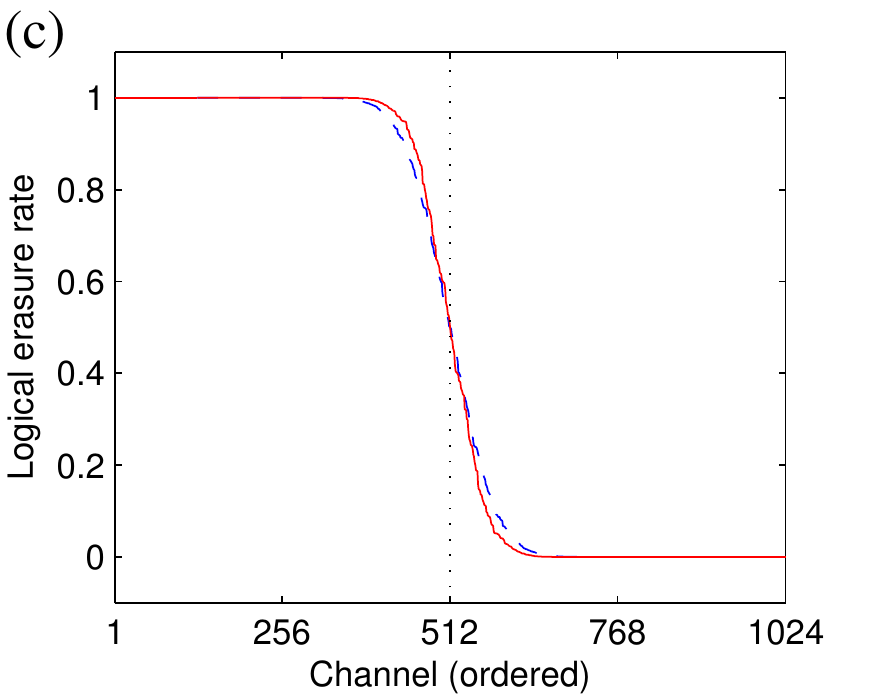}\includegraphics[width=0.5\doublecolumnwidth]{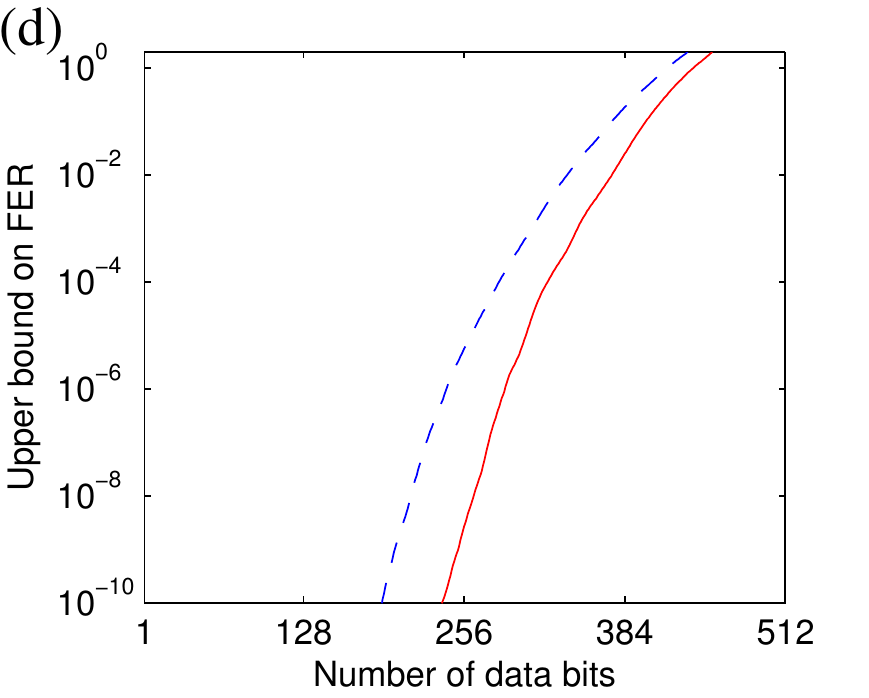}
\caption{Polarization of the logical channels under 1024-bit polar and branching-MERA codes with the 50\% erasure channel. (a,b) The accuracy of each channel, parameterized by effective erasure rate of the logical channels under successive cancellation decoding, for (a) the polar code and (b) the branching-MERA code. (c) The same data is plotted in descending order for the polar (dashed, blue line) and branching-MERA (solid, red line) codes. The dotted vertical line corresponds to the capacity at 50\%. (d) The cumulative sum gives a simple upper bound to the frame-error rate (FER) for the specified number of number of data bits. \label{fig:channels} }
\end{figure}

We have used this fact to study channel polarization of the erasure channel under polar and branching-MERA coding. In \fig{channels}, we present results for the 50\% erasure channel over 1024 bits. In \fig{channels}~(a,b,c) we observe that the branching-MERA contains somewhat fewer channels in the intermediate area between the perfect and useless limits, and further, that the good channels are a little more strongly localized on the left (and conversely the bad channels are localized on the right). This latter fact is particularly significant for successive cancellation decoding because it means more information regarding the frozen bits is available to the decoder when determining the data bits, reducing the gap in performance between maximum likelihood and successive cancellation decoding.

From these results we can also deduce a simple upper bound on the probability of at least one error in the block, or frame-error rate (FER).  By simply summing the probability of erasures over the $k$ data channels and noting that the chance of an error is \emph{at least} the sum of probabilities that a given data bit is the first to be decoded incorrectly, we arrive at an (over)-estimate of the FER. In \fig{channels}~(d) we show this sum for a range of data rates. The upper bound suggests that the branching MERA can deliver a significant increase in the amount of data sent for a fixed error-rate (especially for small target error rates).

As a practicality, logical channels are selected according to their individual rates of success, neglecting any correlations between errors (this should be a reasonable assumption in the low error-rate regime). One can approximate the logical channel capacities for both the polar code and branching-MERA code for arbitrary error-models using Monte Carlo sampling~\cite{A09a} or more sophisticated techniques~\cite{Tal2013}, however we do not investigate this here. In fact, the simulations in the next section are done using a simplified channel selection procedure, detailed below.

As a final note, we have performed a preliminary numerical investigation of the \emph{rate} of channel polarization and have observed (using the upper bound above and the erasure channel) that for a fixed data rate $k/n$ (less than the capacity) both codes have frame-error rates scaling like $\exp(-c\sqrt{n})$. However the constant $c$ is larger for the branching-MERA code than the polar code. It would be interesting to study how the FER varies with data rate $k/n$ and the capacity in both cases. 

\section{Numerical simulation of error-correction capabilities}
\label{sec:results}

Here we numerically compare the performance of the polar and branching MERA codes at protecting data from a variety of channels, focussing on finite-code length effects on codes between 256 and 8192 bits. For all our simulations we have used a simplified channel selection scheme that is independent of the details of the error model. This works by simply performing a `dry run' of the decoding algorithm using an all-zero input and an output where the decoder believes each bit has an independent probability $p$ of being a 1, and $1-p$ of being a 0. For each bit $x_i$ we contract the corresponding tensor-network diagram (with 0-valued bits to the right, i.e. $x_j = 0$ for $j > i$) and use the resulting probability distribution $P(x_i)$ as a measure of the logical channel fidelity. We have found that this procedure is simple to implement and gives adequate results over all the channels we have investigated (for instance, performing better for the bit-flip channel than the data presented in \fig{channels}, which derives from the erasure channel). A slight improvement in performance can be observed by using the channel selection tailored to the specific error model in question, but the comparative performance between the polar and branching-MERA codes remains very similar, and this procedure is adequate for the purposes of this section.

\begin{figure}[t]
\centering
\includegraphics[width=0.5\doublecolumnwidth]{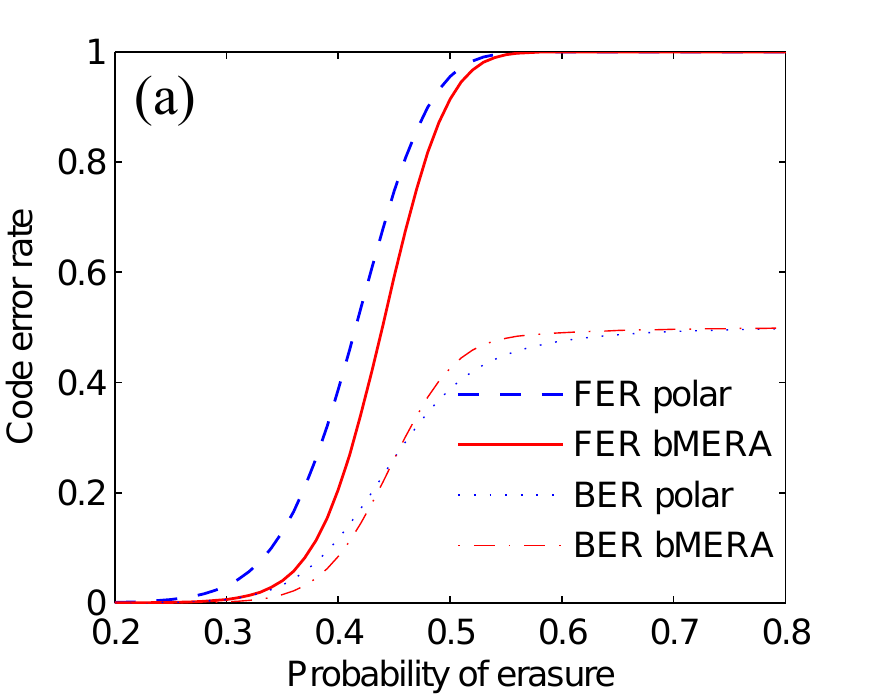}\includegraphics[width=0.5\doublecolumnwidth]{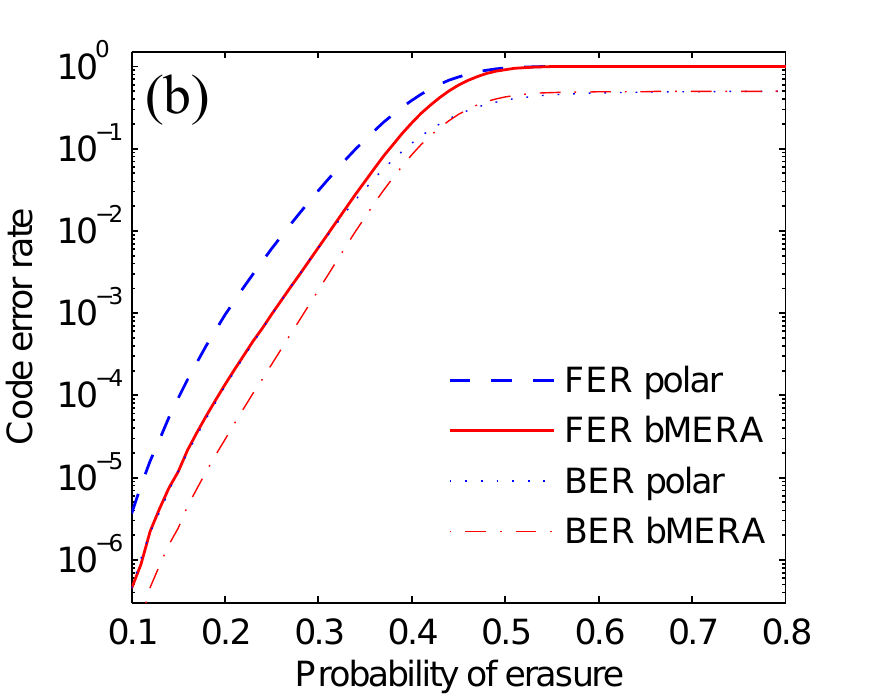}
\includegraphics[width=0.5\doublecolumnwidth]{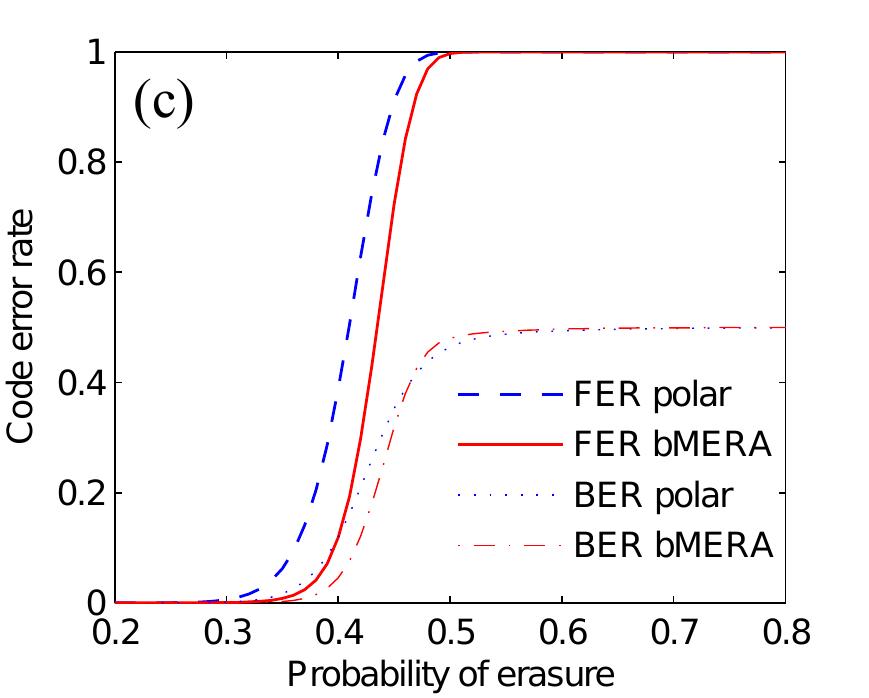}\includegraphics[width=0.5\doublecolumnwidth]{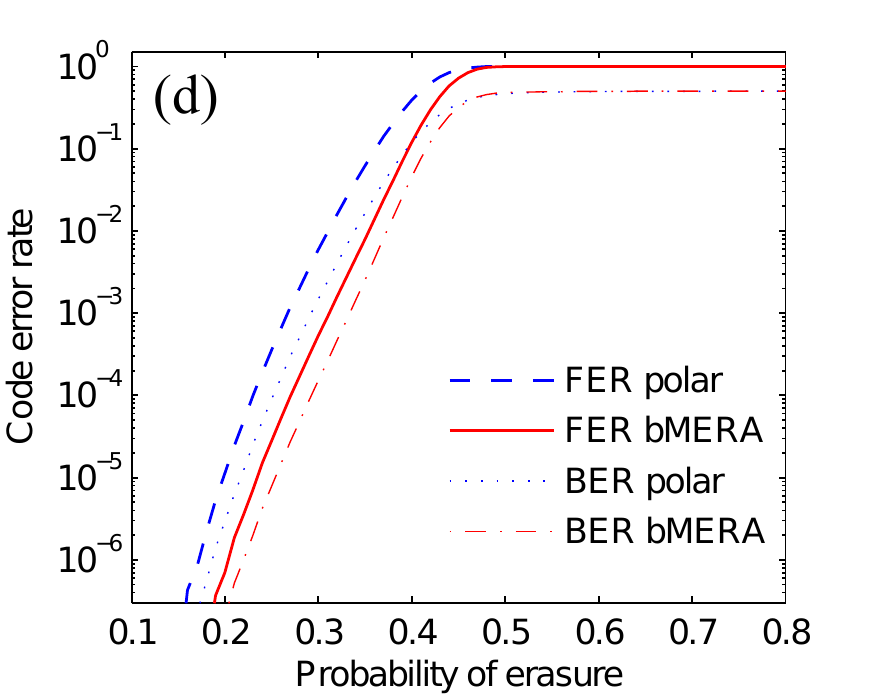}
\includegraphics[width=0.5\doublecolumnwidth]{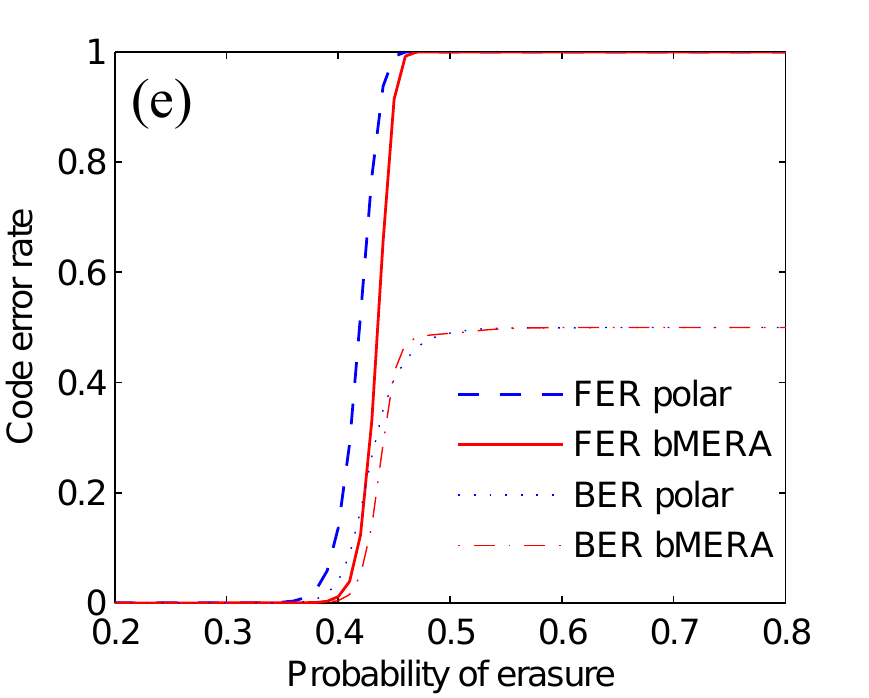}\includegraphics[width=0.5\doublecolumnwidth]{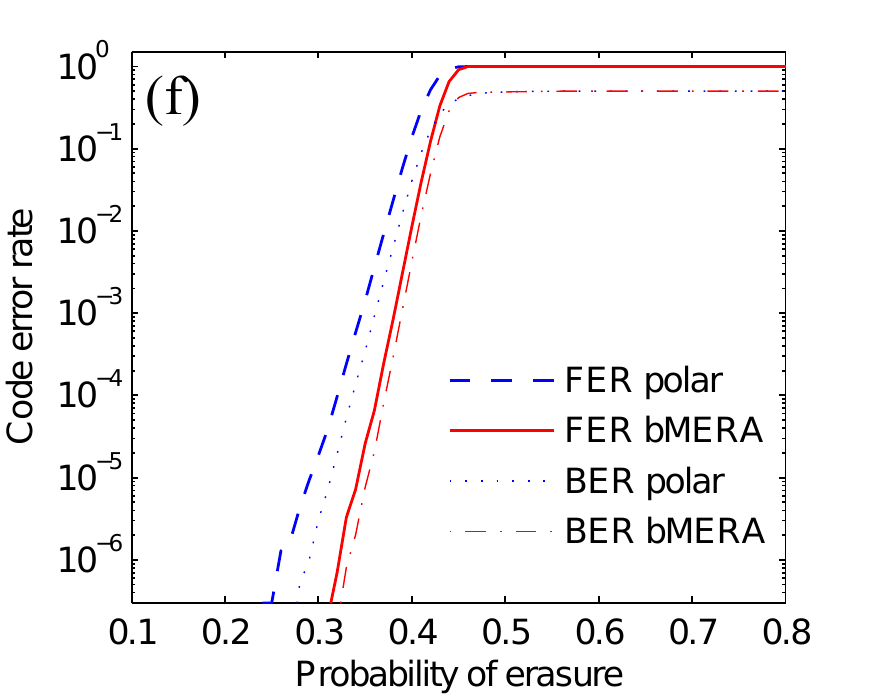}
\caption{Comparison of the performance of rate 1/2 polar and branching MERA codes of various sizes for the binary erasure channel. The encoded message contains (a,b) 256 bits. (c,d) 1024 bits and (e,f) 8192 bits. The capacity with erasure probability 0.5 corresponds to the code rate 1/2. \label{fig:classical_erasure} }
\end{figure}

The results for the binary erasure channel with code-rate 1/2 are given in \fig{classical_erasure}. In all cases we observe several things. Finite-size effects are significant in both codes, with the waterfall region separating ``perfect" and ``useless" behavior being somewhat below the capacity of the erasure channel (which suggests that erasure rates of up to 0.5 are tolerable for our encoding rate). Nonetheless, the threshold of the branching MERA code is significantly closer to this value than the polar code. On a logarithmic scale, is it evident that the performance in the low-error region is significantly better --- note the slope in \fig{classical_erasure}~(f) is significantly greater for the branching MERA code. Neither code displays any evidence of an error floor (nor is it expected). Finally, both codes display a tendency for any error to be catastrophic --- involving errors on many bits. The ratio between the bit error rate (BER) and frame error rate (FER) is very large for the polar code and even higher (close to 0.5) for the branching MERA code. This corresponds to either a perfectly decoded message or a completely scrambled one. Interestingly, this is the behavior expected of a ``perfect" random code as Shannon envisaged, where the most likely messages are completely uncorrelated. It would be worth investigating whether techniques such as systematic encoding could be used to reduce the BER and, in particular, reduce the number of incorrect bits to a low enough level that a second layer of encoding may be of practical use.

\begin{figure}[t]
\centering
\includegraphics[width=0.5\doublecolumnwidth]{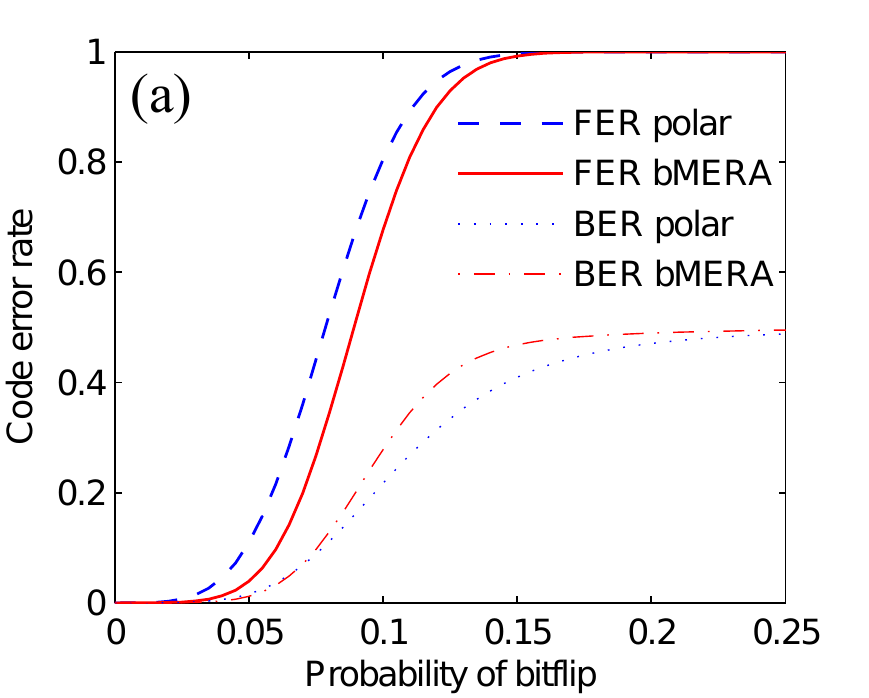}\includegraphics[width=0.5\doublecolumnwidth]{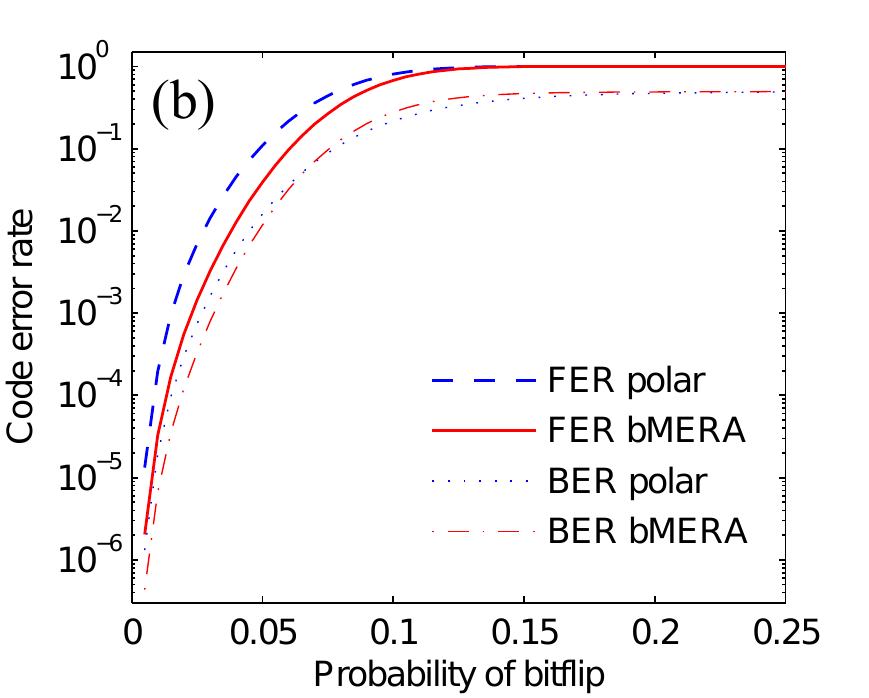}
\includegraphics[width=0.5\doublecolumnwidth]{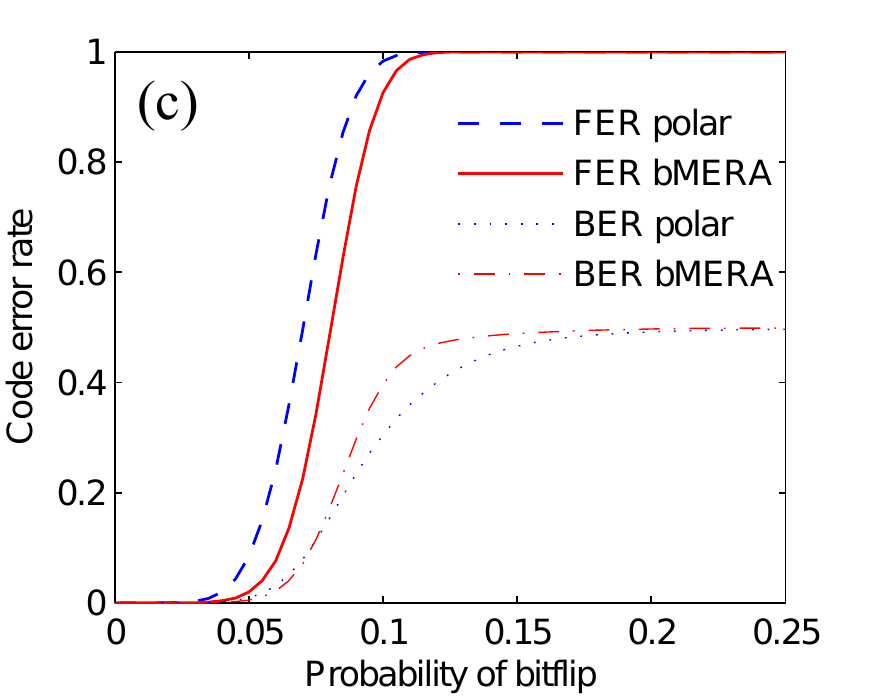}\includegraphics[width=0.5\doublecolumnwidth]{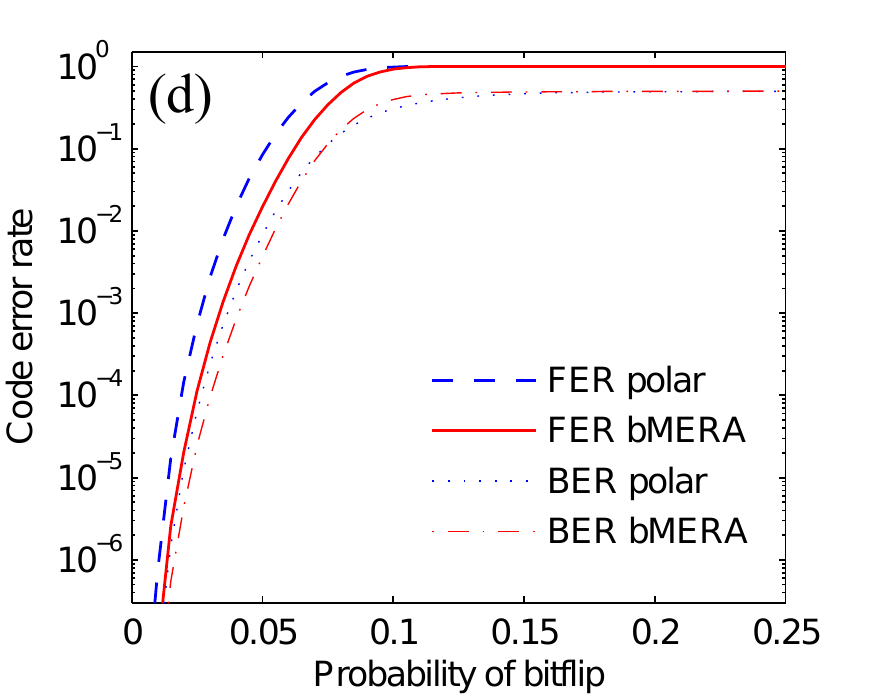}
\includegraphics[width=0.5\doublecolumnwidth]{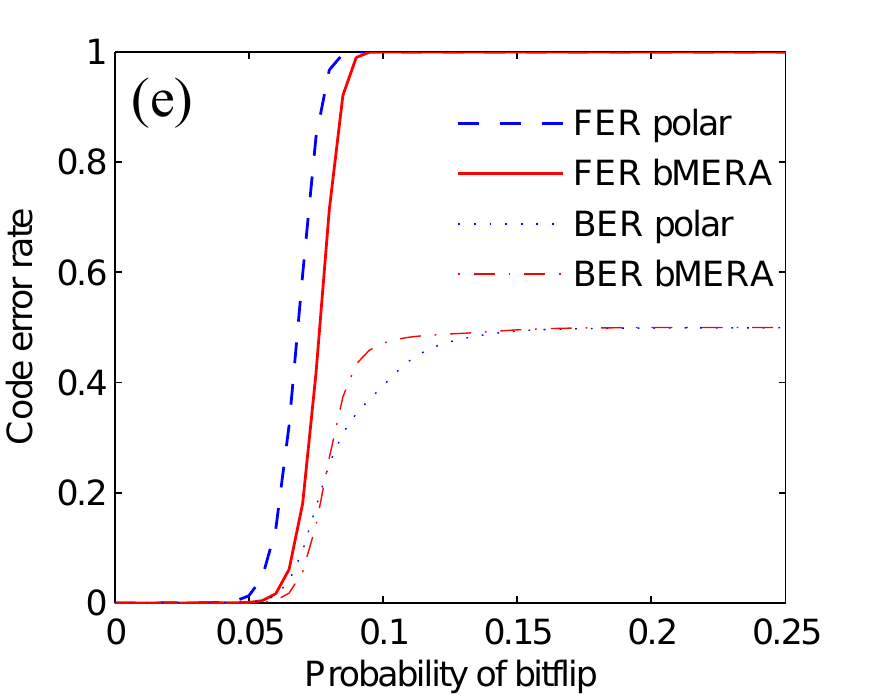}\includegraphics[width=0.5\doublecolumnwidth]{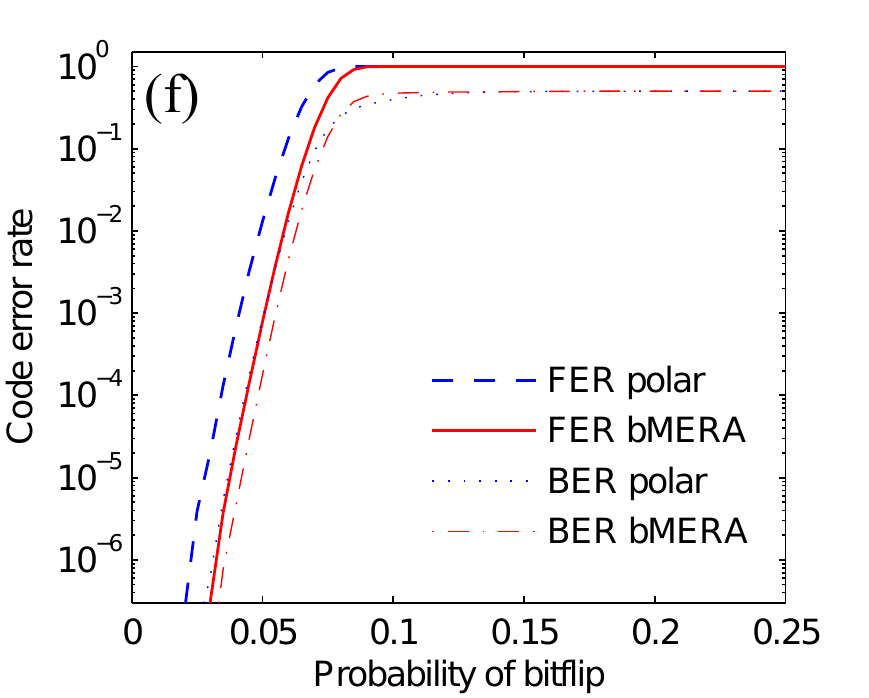}
\caption{Comparison of the performance of rate 1/2 polar and branching MERA codes of various sizes for the bit-flip channel. The encoded message contains (a,b) 256 bits. (c,d) 1024 bits and (e,f) 8192 bits. The capacity with bit-flip probability approximately 0.11 corresponds to the code rate 1/2. \label{fig:classical_bitflip} }
\end{figure}

In \fig{classical_bitflip} we see similar behavior for the bit-flip channel. In this case we observe even greater finite-size effects, with the observed waterfall regions quite a bit below the expected threshold at a bit-flip rate of approximately 0.11. The branching MERA code performs better in all cases, with a higher tolerance for error, a sharper transition between good and bad performance, and better scaling in the low error-rate region.

\begin{figure}[t]
\centering
\includegraphics[width=0.5\doublecolumnwidth]{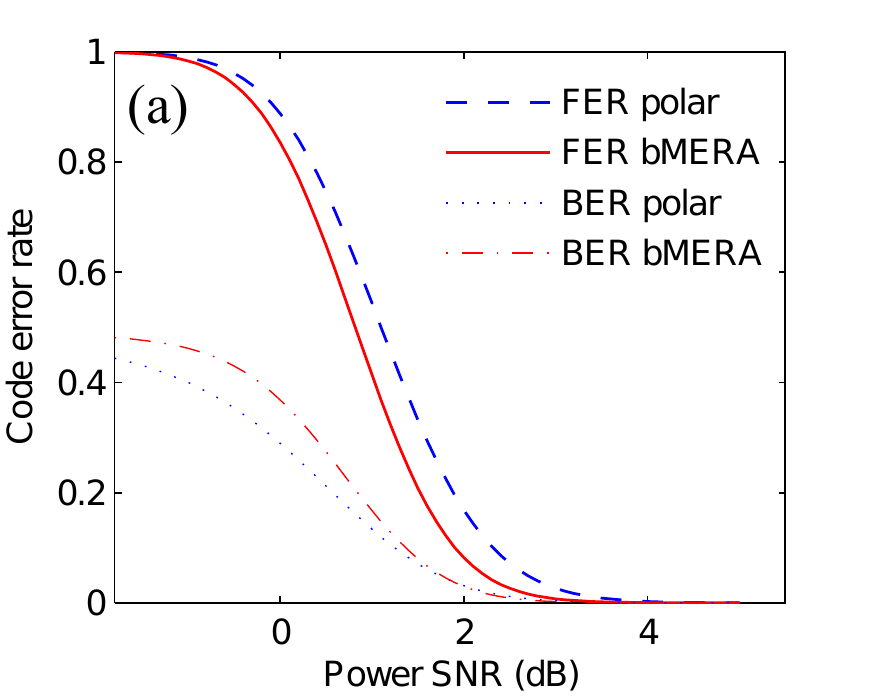}\includegraphics[width=0.5\doublecolumnwidth]{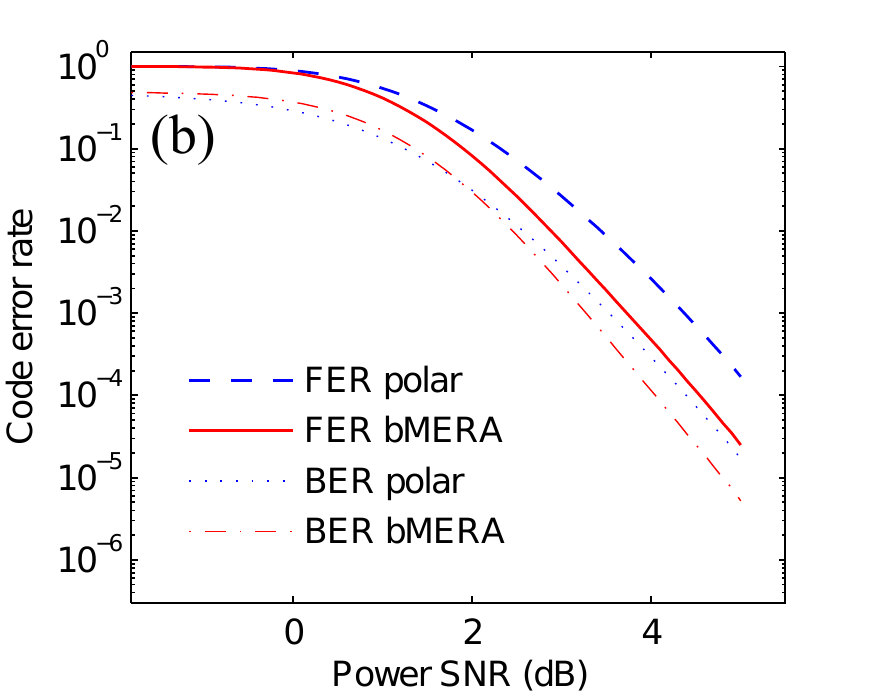}
\includegraphics[width=0.5\doublecolumnwidth]{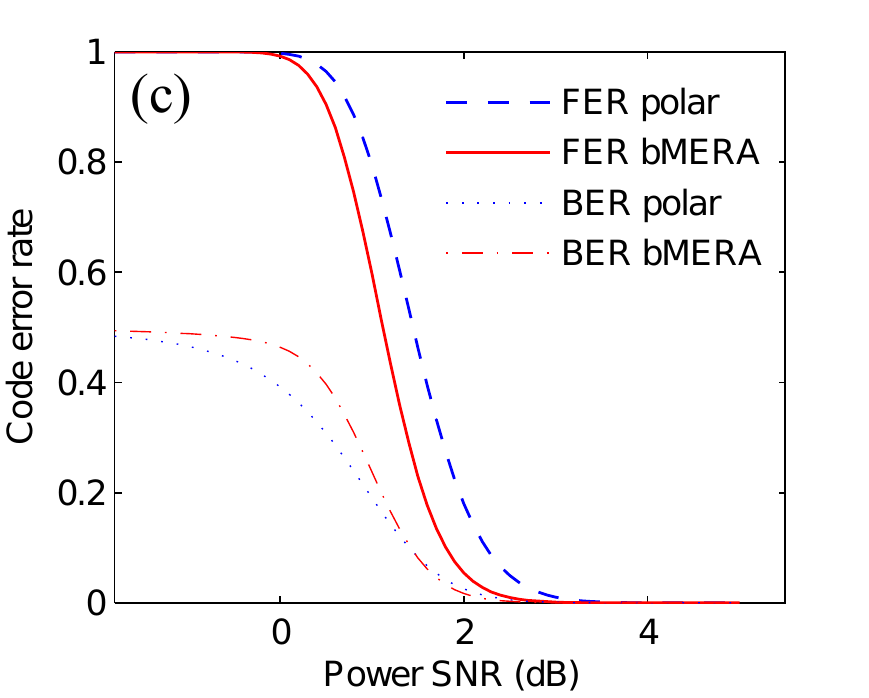}\includegraphics[width=0.5\doublecolumnwidth]{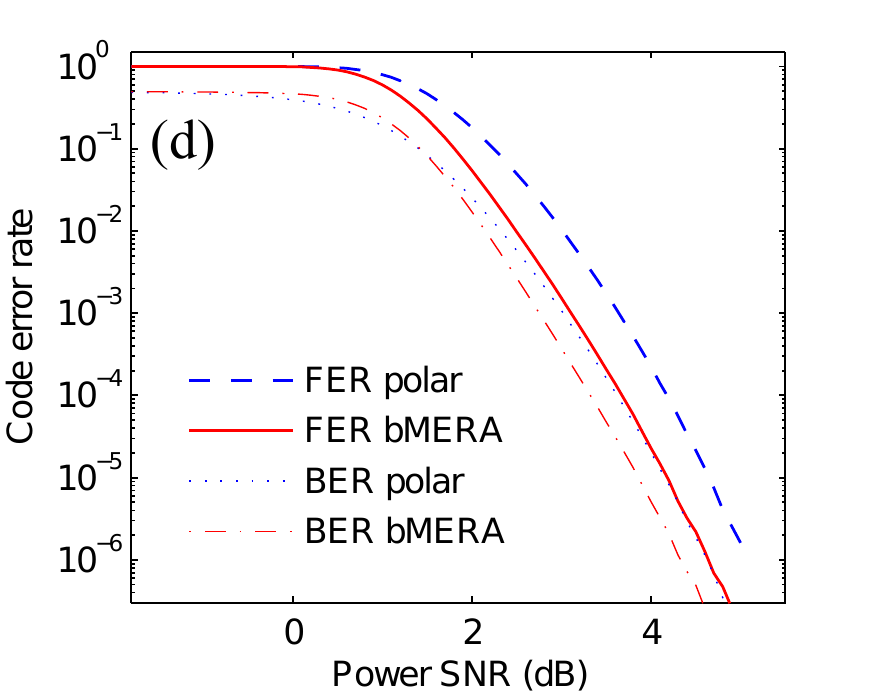}
\includegraphics[width=0.5\doublecolumnwidth]{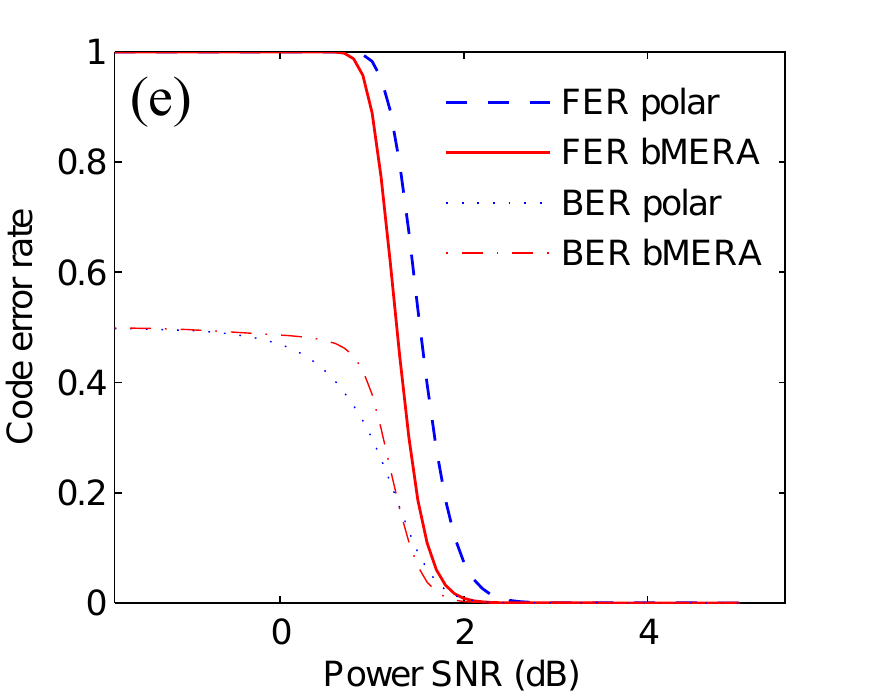}\includegraphics[width=0.5\doublecolumnwidth]{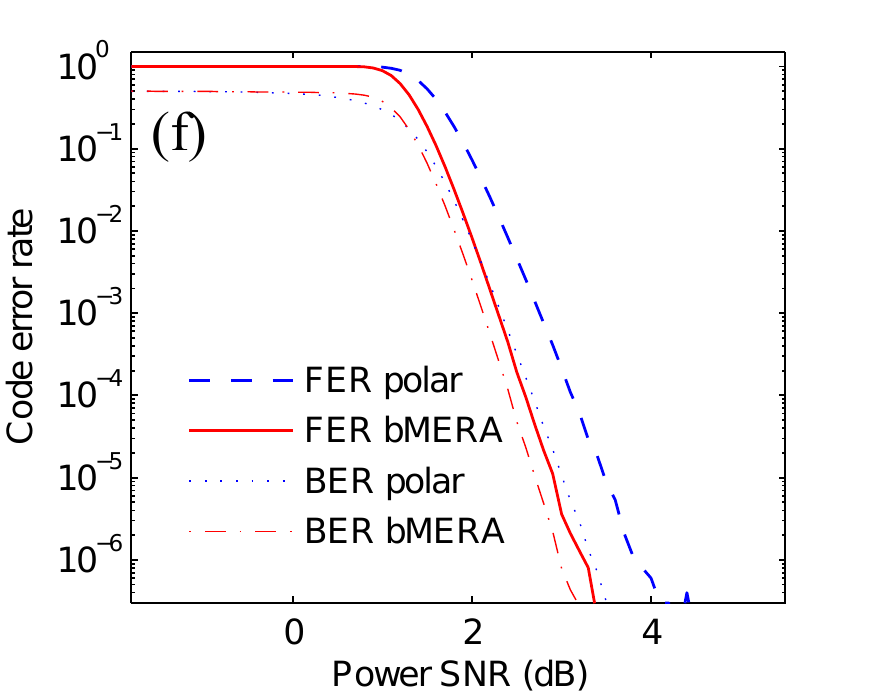}
\caption{Comparison of the performance of rate 1/2 polar and branching MERA codes of various sizes for the additive Gaussian white noise channel. The encoded message contains (a,b) 256 bits. (c,d) 1024 bits and (e,f) 8192 bits. \label{fig:classical_agwn} }
\end{figure}

Finally, we investigated performance under the more realistic additive Gaussian white noise channel in \fig{classical_agwn}. Once again we observe similar behavior: the branching MERA code has better error performance than the polar code, including tolerance for larger noise rates and better scaling in the low noise region.

Based on these results, we can conclude that the branching MERA code is a significant improvement to the polar code when it comes to error correction capabilities. Also, the numerical cost is not changed significantly, with the same scaling and, anecdotally, approximately twice the computation effort to decode.

\section{Conclusion and Discussion}
\label{sec:discussion}

We have demonstrated a general connection between the problem of decoding an error correcting code and the graphical calculus of tensor networks used in the field of quantum many-body physics. Using a family of tensor network recently introduced in that setting, we presented  a new family of error-correcting codes that generalize polar codes in a  natural way. Recasting the decoding problem as a tensor network contraction, we have demonstrated that Arikan's sequential decoder can be realized with log-linear complexity,  requiring roughly twice the computational effort of polar codes sequential decoding. 

Our numerical simulations show that this new code outperforms polar codes in several ways, including stronger channel polarization and enhanced error-correcting performances.
On the other hand, there clearly is more room for improvement, so that finite-size performance is closer to capacity. For instance, more complex schemes for channel selection may be possible. We have performed an additional analysis of the maximum likelihood decoder for smaller polar and branching MERA codes under the erasure channel and our results indicate performance significantly closer to capacity than observed in \fig{classical_erasure}. This difference arises because at all stages of decoding, every syndrome measurement is available to be used, unlike the successive cancelation decoder which only has access to previous bits. We speculate that the main advantage of branching MERA code compared to the polar code is that the syndrome bits are more tightly clustered to one side and the data channels on the other --- thus increasing the information available to the successive cancellation decoder for the earlier data bits. 

The connection between tensor networks and coding opens the door to many other encoding schemes. Only within the family of branching MERA networks, many different codes can be obtained by varying the elementary gates in the network and increasing the number of bits in elementary gates (i.e. increasing the ``bond dimension" in the TN language). Other tensor networks could also be considered along with their heuristic contraction schemes, e.g. \cite{VC04a,GLW08a}. In a similar vein, other decoders including belief propagation~\cite{Eslami2010} and list decoding~\cite{Tal2011} could also enhance the error-correction performances. Lastly, quantum versions of these codes can also be defined and similarly outperform quantum polar codes \cite{FP13}.

\section*{Acknowledgments}
The authors would like to Jean-Pierre Tillich for useful discussions. AJF would like to thank TOQATA (Spanish grant PHY008-00784), the EU IP SIQS, and the MPI-ICFO collaboration for supporting this research. DP would like to acknowledge support from NSERC and FQRNT through the network INTRIQ. Computational resources were provided by Compute Canada and Calcul Quebec.

\bibliographystyle{IEEEtran}
\bibliography{qubib}

\end{document}